\documentclass{PoS}

\title{Spherocity study for $e^{+}e^{-}$  Fragmentation Functions using Pythia MC generator}

\ShortTitle{Short Title for header}

\author{\speaker{H. Bello Mart\'inez}, R. J. Hern\'andez-Pinto, I. Le\'on Monz\'on \\
       Facultad de Ciencias F\'isico-Matem\'aticas, Universidad Aut\'onoma de Sinaloa,\\
       Ciudad Universitaria, CP 80000, Culiac\'an, Sinaloa, M\'exico.\\
        E-mail: \email{hectorbm884@gmail.com, roger@uas.edu.mx, \\ ildefonso.leon.monzon@cern.ch}}

\abstract{The Fragmentation Functions is one of the non-perturbative components of the QCD factorization theorem. They represents the probability of a parton carrying a fraction $z$ of momentum to form into a particular kind of hadron. In this work, we study the jet fragmentation functions in the collisions between electrons and positrons. The jets where identified with Fastjet for different $p^{ch jet}_{T}$ intervals. The intervals and the final jets were reconstructed by means of the event shape separation using spherocity variable, the study is performed under Pythia Monte Carlo event generator framework.}

\FullConference{7th Annual Conference on Large Hadron Collider Physics - LHCP2019\\
		20-25 May, 2019\\
		Puebla, Mexico}

\begin{document}

\section{Introduction}
\footnote{This contribution is based on our work \cite{HRI} } In recent years the Large Hadron Collider (LHC) has obtained very interesting results in the study of systems created in pp collisions at center of mass energy of the order of TeV. In addition, past results at CERN for $e^{+}e^{-}$ have motivated to start new studies in new accelerator facilities such as the Future Circular Collider (FCC) at greater energies \cite{FCC1,FCC2,FCC3,FCC4}, Furthermore, there is an increasing interest to make studies in these two colliding systems, due to the fact that the FCC will collide not only protons but electrons and positrons. The study of the final states particles produced after the collision gives useful information of the hadronization process \cite{Alb}. In particular, the study of fragmentation functions (FF) helps to understand and should model the theory behind this non-perturbative phenomena \cite{Florian,Tana}. On the other hand, spherocity analysis \cite{Banfi} is a tool that has taken a great interest  for studying the event shapes of final states charged particles \cite{Alice1} and hadrons \cite{Gyula,Vita}. Important results had been obtained by computing the production of reconstructed jets with Fastjet \cite{Fastjet} after the hadronization process \cite{HAA,HBM,Alice2,Alice3}. There have been recent studies in which they consider to join these topics together at generator level using Pythia \cite{Py6,Py8,Monash}, in order to extract a more differential analysis of the final states created in $e^{+}e^{-}$ collisions.

In this work, we study the fragmentation functions for jet production using two different approaches for jet recconstruction, the anti-$k_{T}$ algorithm and the spherocity analysis, this was done using Pythia 6 and 8 Monte Carlo event generators.


\subsection{ Fragmentation Functions}

The final stage hadron, $d\sigma/dz$, in general is described by mean of the QCD factorization theorem \cite{Alb}. It allows us to compute observables in terms of short range quantities (calculable in perturbative QCD) and long range quantities (not calculable in perturbative QCD), which are momentum fraction $z$ dependent. The fragmentation functions are defined as follows:



\begin{itemize}
 
\item{ For electron positron annhilation,} due to the lack of internal structure of electrons in the procces $e^{+}e^{-} \rightarrow \gamma/Z^{0} \rightarrow h + X$, the hadro-production cross section leads to the description of the fragmentation function $F^{h}(z,s) $ as \cite{Tana}: 

\begin{eqnarray}
\frac{1}{\sigma_{0}}\frac{d\sigma^{h}}{dz}=F^{h}(z,s)=\sum_{i}\int^{1}_{z}\frac{dz}{z}C_{i}(z,\alpha_{s},\frac{s}{\mu^{2}})D^{h}_{i}(z,\mu^{2})+\mathcal{O}(1/\sqrt{s}),
\end{eqnarray}

here $\sigma_{0}$ is a normalization factor, $i$ runs for all quarks and gluons, $D^{h}_{i} (z,\mu^{2})$ are the fragmentation densities that describes the probability that the parton $i$ fragments into a hadron h. the $C_i$ are the observable-dependent coefficient functions whose NLO and NNLO terms are known \cite{NLO,NNLO}, and
$z =\frac{2p_{h}\cdot q}{q^{2}}$. In practice the approximations $z \approx 2P_{h} /\sqrt{s}$ and $z \approx 2E_{h} /\sqrt{s}$ are used.





\item{ For jets,} the jet production cross section is given by 

\begin{eqnarray}
\frac{d\sigma^{ch jet}}{dp_{T}d\eta} (􏰈p^{ch jet}_{T})􏰉 = \frac{1}{\mathcal{L}^{int}} \frac{\Delta N_{jets}}{\Delta p_{T}\Delta\eta} (􏰈p^{ch jet}_{T}), 
\end{eqnarray}

where it has been defined the scaling variable $z^{ch} = p^{particle}_{T}/p^{ch jet}_{T}$,  and the jet fragmentation function is defined \cite{Alice3} as: 
   
\begin{eqnarray}
F^{z}(z^{ch},p^{ch jet}_{T})=\frac{1}{N_{jets}}\frac{dN_{jets}}{dz^{ch}}.
\end{eqnarray}
   
\end{itemize}

It is worth mentinoning that a precise knowledge on FFs is vital for the quantitative description of a wide variety of hard scattering processes and the understanding of phenomena produced by jets \cite{Alice1,HBM,Alice2,Alice3}.
   
 \subsection{ Jet reconstruction methods, anti-$k_{T}$ algorithm and spherocity}

One of the implemented algorithm in Fastjet  \cite{Fastjet} for jet reconstruction is the anti-$k_{T}$ \cite{Antikt}. It consists in taking the beam jet axis $B$, and to select all the particles $i$ with distance $d_{iB} = 1/p^{2}_{Ti}$, inside the radius $\Delta R^{2}_{ij} = (y_{i} -y_{j})^{2} +(\phi_{i} -\phi_{j})^{2}$. The distance among each particles pair $i$, $j$ is given by,

\begin{eqnarray}
 d_{ij} = min(1/p^{2}_{Ti},1/p^{2}_{Tj} ) \Delta R^{2}_{ij} /R^{2}.
 \end{eqnarray}
 
\begin{figure}
\begin{center}
  \includegraphics[width=0.7\linewidth]{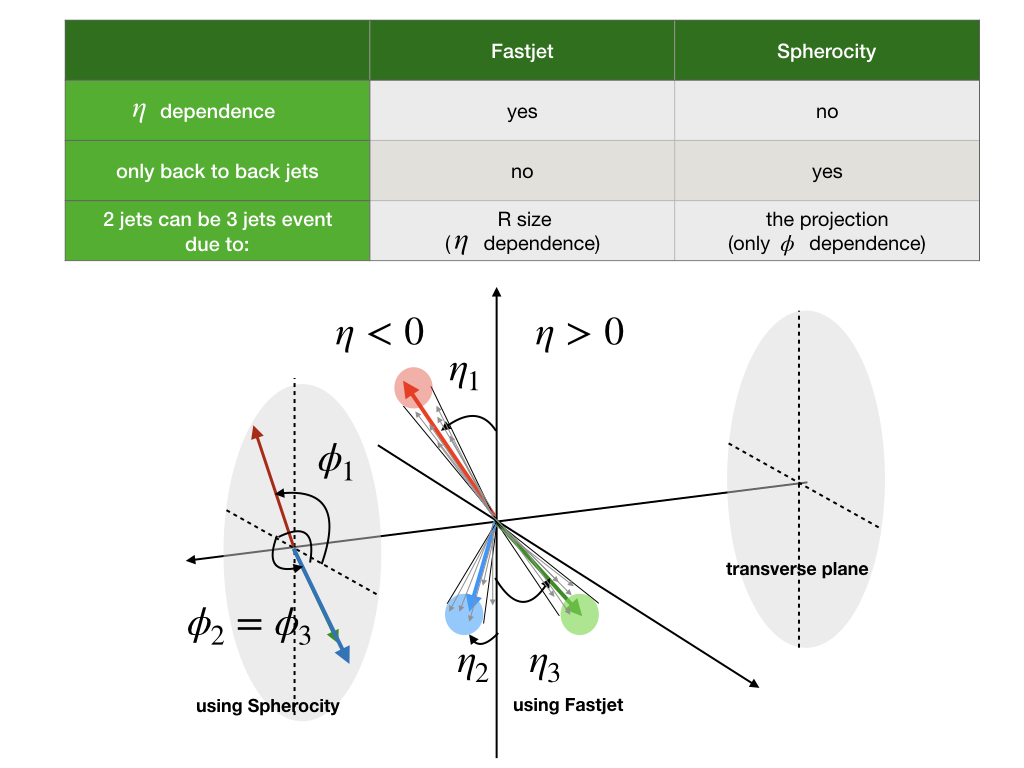}
  \caption{Graphical description of the two different methods for jet reconstruction explaining the complementarity of the methods.}
  \label{fig:FJSO}
  \end{center}
\end{figure}

On the other hand, the event shape called transverse spherocity \cite{Banfi} is defined as,

\begin{eqnarray}
S^{pherocity}_{T}=S_{O}=\frac{\pi^{2}}{4} \min\limits_{ \vec{n}=(n_{x},n_{y},0)} \left( \frac{\sum_{i} |\vec{p}_{Ti}\times \vec{n} |}{\sum_{i} | \vec{p}_{Ti}|} \right)^{2},
\end{eqnarray}

where the values which defines the structure in the transverse plane runs from 0 (for jetty) to 1 (for isotropic). It is remarkable that the two methods are essentially different and their use should provide complementary information. This naive conclusion is based on the fact that in Fastjet we have an $\eta$ dependence due to the selection of radius of the cone but in spherocity there is only $\phi$ dependence. Furthermore, with spherocity only back to back dijet shapes can be selected, meanwhile in Fastjet this feature is not limited, see Figure \ref{fig:FJSO}.

\newpage

\subsection{Fragmentation Functions in Pythia}

The FF inside Pythia 6.4 \cite{Py6} and Pythia 8.2 \cite{Py8} is basically given by the Lund  symmetric FF as \cite{Lundff}: 

\begin{eqnarray}
f(z)\propto z^{-1}(1-z)^{a}exp\left(\frac{-bm^{2}_{\perp}}{z}\right),
\end{eqnarray}

where $m_{\perp}^2$  is the hadron transverse mass squared, 
$z$ is the fraction hadron to parton momentum 
and the free parameters  $a$ and $b$ are related to the total multiplicity; 
these parameters govern the shape of the FF, and must be constrained by fits to data, Roughly speaking, large value for $a$ suppress the hard region $z \rightarrow 1$, while a large value for b  suppresses the soft region, $z \rightarrow 0$. For Pythia 6.4 \cite{Py6} these values are setted to $a=0.3$, $b=0.58$; and Pythia 8.2 is using the Monash tune \cite{Alice3}, where the default values are, $a=0.68$, and $b=0.98$. 


\section{Event selection}

For the simulation, we generate 30 million of events in Pythia 6.4 \cite{Py6}   and Pythia 8.2 \cite{Py8} using tune Monash \cite{Monash} for 
$e^{+}e^{-}$ collisions at $\sqrt{s} = 91$ GeV, using DELPHI kinematic cuts \cite{DELPHI}, at least 5 stable charged particles within 0.3 $< p < $70 GeV, E$ > 15$ GeV and $0.01 < z < 1$.


For jet reconstruction, the anti-$k_{T}$ algorithm was used with the kinematical cuts $p_{T,min} = 5$ GeV, 
$\eta_{jet} < 0.5$ and $R=0.4$, and finally we use the $p_{T}$ recombination scheme as in ALICE method \cite{Alice3}. For the spherocity study, the percentile selection ($S_{Opc}$) was taken in intervals of $10\%$  as it was done in Ref. \cite{Alice1}.

\section{Results}

\begin{figure}
\begin{center}
  \includegraphics[width=\linewidth]{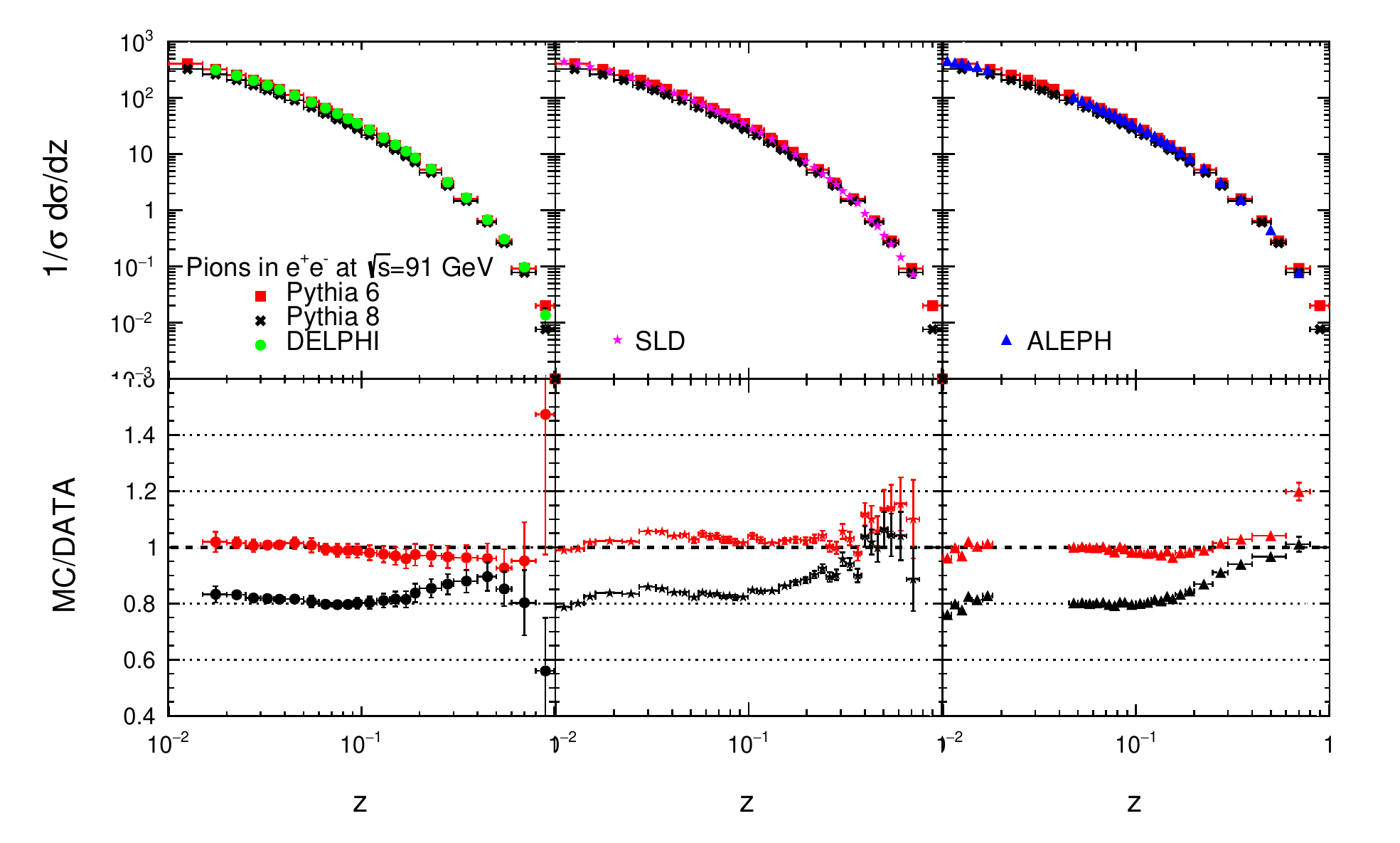}
   \vspace{-0.5cm}
  \caption{On top, the  pions FFs (top) for $e{+}e^{-}$ at $\sqrt{s} = 91$ GeV. Pythia 6.4, Pythia 8.2, DELPHI \cite{DELPHI}, SLD \cite{SLD} and ALEPH \cite{ALEPH} data are shown. On bottom, the comparison data to Monte Carlo. 
  }
  \label{fig:FFPID}
  \end{center}
\end{figure}

\begin{figure}
\begin{center}
  \includegraphics[width=\linewidth]{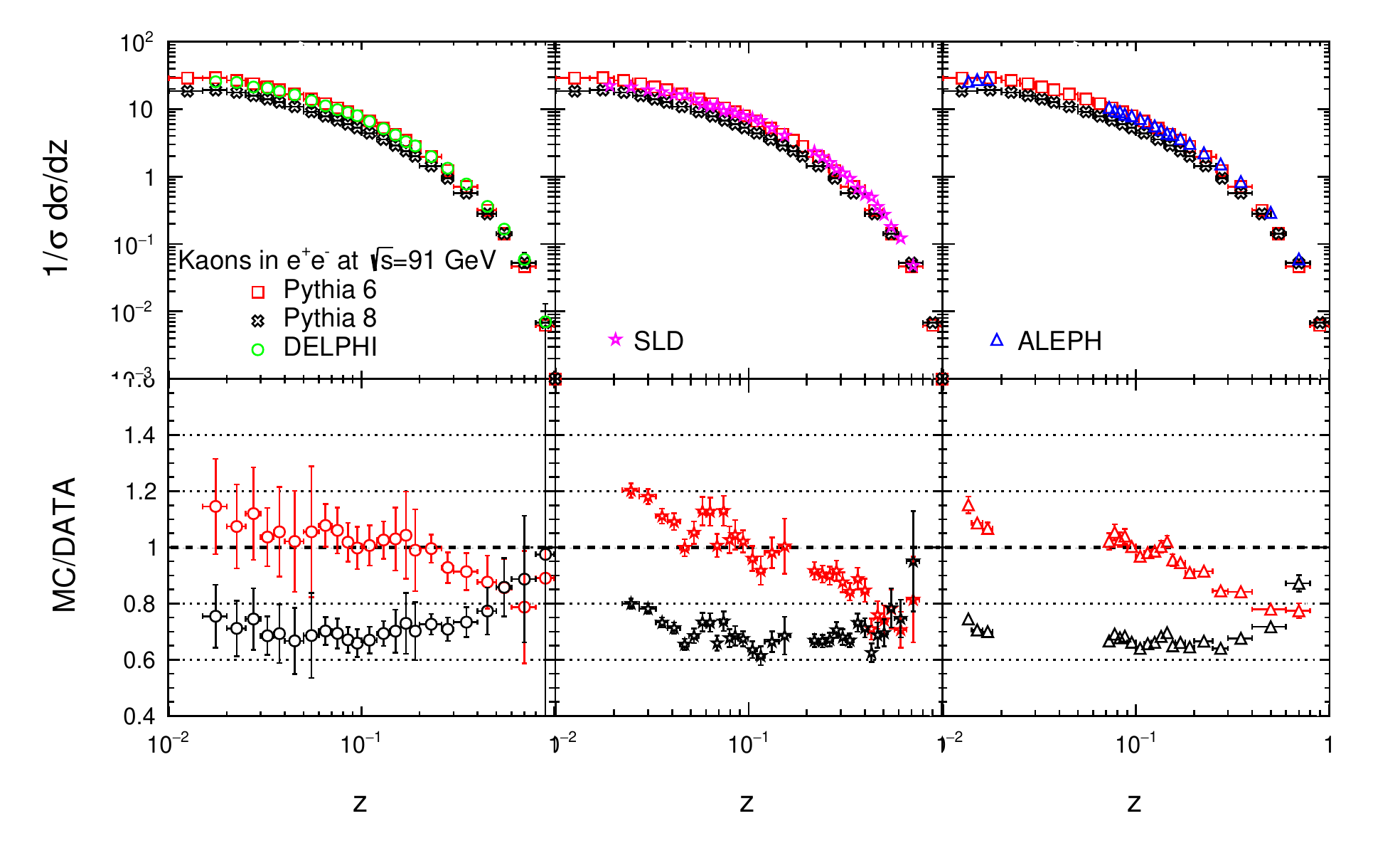}
    \vspace{-0.5cm}
  \caption{On top, the  kaons FFs (top) for $e{+}e^{-}$ at $\sqrt{s} = 91$ GeV. Pythia 6.4, Pythia 8.2, DELPHI \cite{DELPHI}, SLD \cite{SLD} and ALEPH \cite{ALEPH} data are shown. On bottom, the comparison data to Monte Carlo. 
  }
  \label{fig:FFPID2}
  \end{center}
\end{figure}

First we present the FF for pions and kaons produced in $e^{+}e^{-}$ collisions at $\sqrt{s} = 91$ GeV, in Figure  \ref{fig:FFPID} and Figure  \ref{fig:FFPID2}, we compare our results with available data from DELPHI \cite{DELPHI}, SLD \cite{SLD} and ALEPH \cite{ALEPH}. For the pions case (Figure  \ref{fig:FFPID})  the difference is within $5\%$ (in Pythia 6.4) to $20\%$ (in Pythia 8.2).
For the kaons case (Figure  \ref{fig:FFPID2})  the difference goes from $20\%$ (in Pythia 6.4) to $40\%$ (in Pythia 8.2), which is expected due to the poor determination of strangeness production in Pythia. It is important to notice that the best agreement is given for the comparison with DELPHI also with the wide binning range of the available data.


Based on these results, we proceed with the analysis for jets reconstructed with FastJet and the event shape characterization with spherocity.

\newpage

\subsection{Jets with Fastjet}

\begin{figure} 
\begin{center}
  \includegraphics[width=0.5\linewidth]{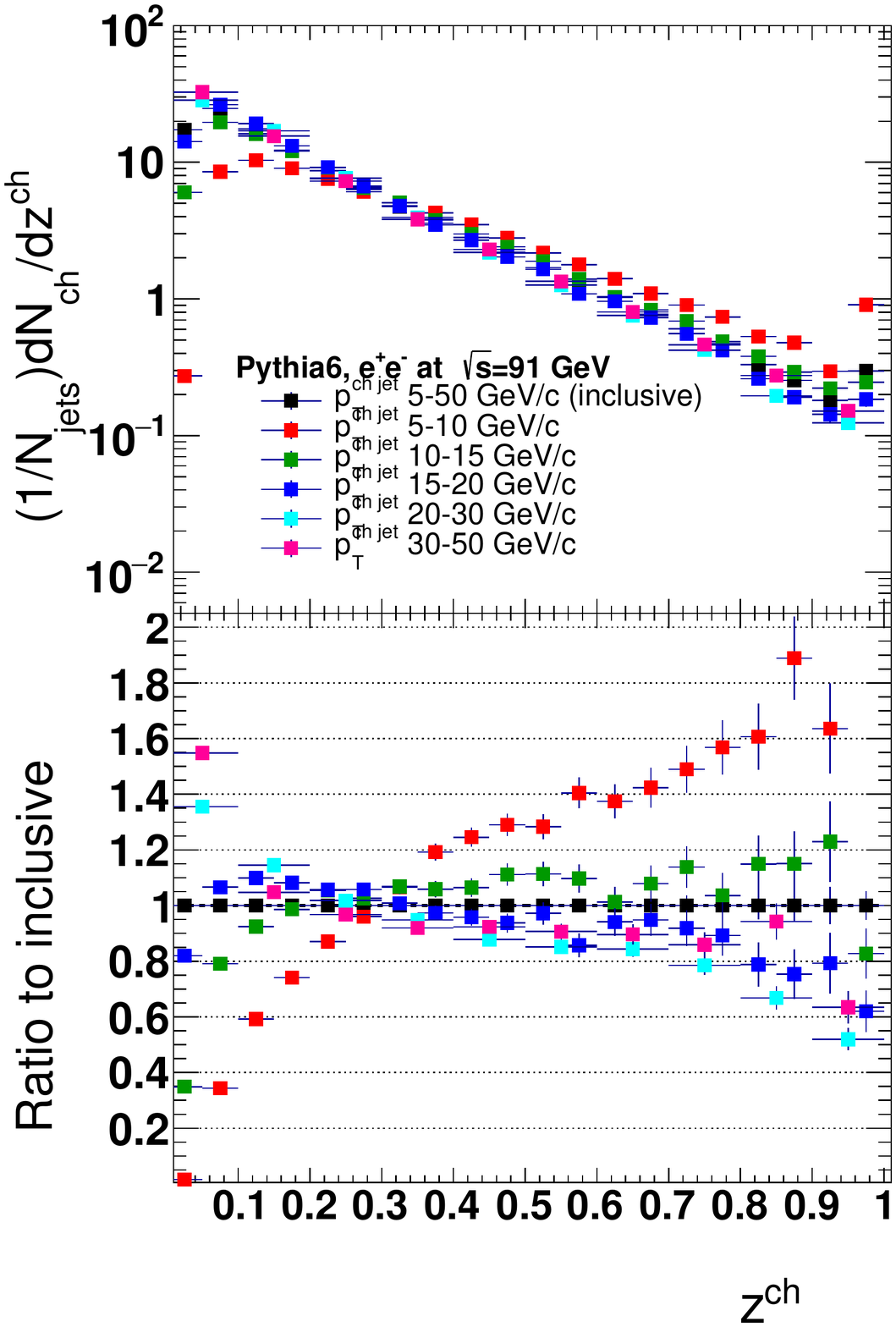}
\vspace{-0.5cm}
  \caption{On top, charged jet fragmentation function, results for $e^{+}e^{-}$ collisions 
  are shown for $p^{ch jet}_{T}$ selection, from the lower (red) to higher (magenta) intervals. On bottom, ratio to inclusive is shown. 
  }
  \label{fig:FFFJ}
  \end{center}
\end{figure}

We present in Figure \ref{fig:FFFJ} the results 
for jets reconstructed in different jet transverse momentum intervals. 
As expected, we obtain large probablities in the regime of high $z$ and low $p^{ch jet}_{T}$, this behaviour is opposite in the small $z$ region. Furthermore, a crossing point was found around $z = 0.3$. A comparison with the inclusive sample (no $p^{ch jet}_{T}$ separation) shows clearly the behaviour. 

\newpage

\subsection{Spherocity separation}

\begin{figure} 
\begin{center}
  \includegraphics[width=0.48\linewidth]{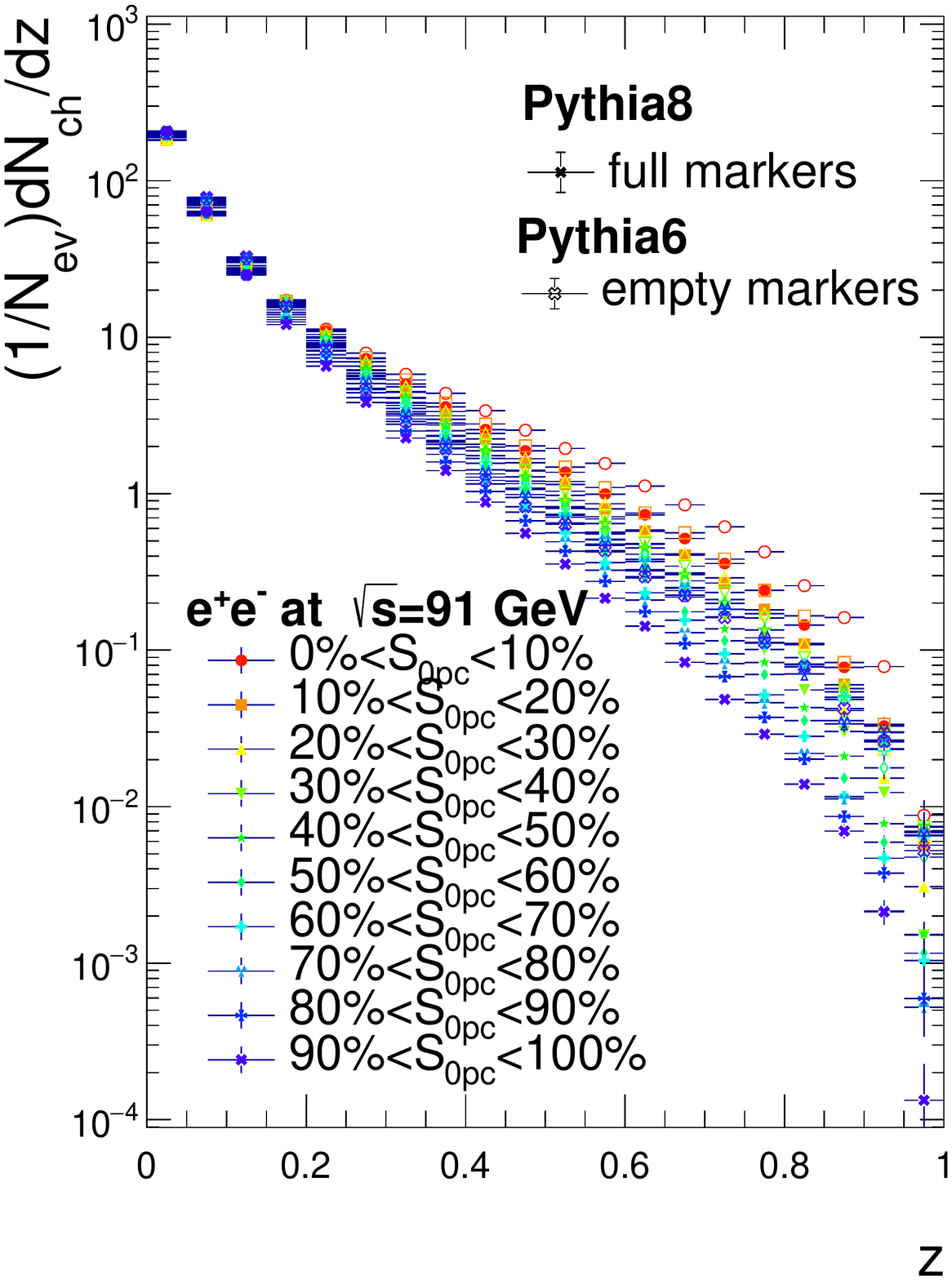}
  \includegraphics[width=0.48\linewidth]{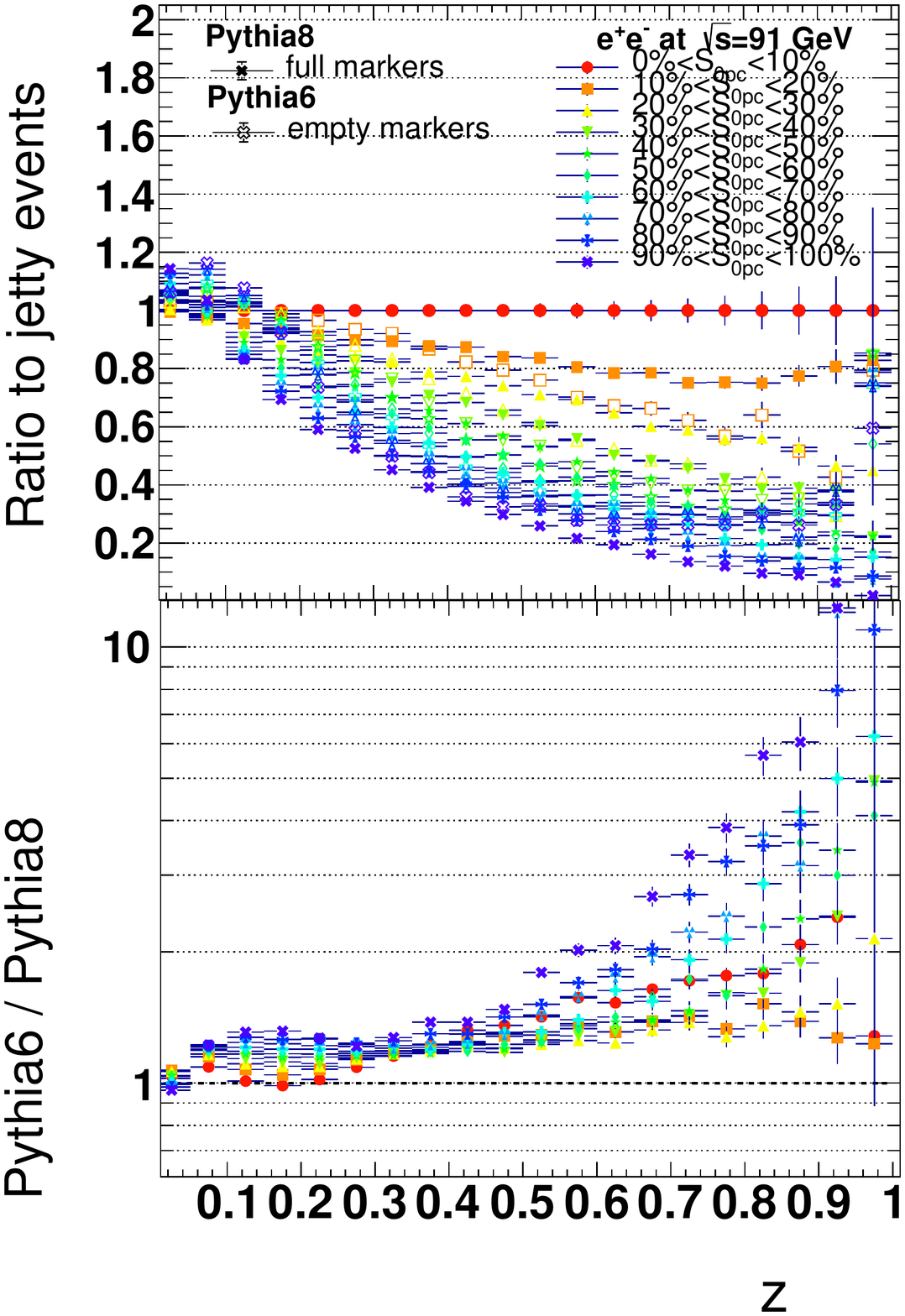}
\vspace{-0.5cm}
  \caption{Left-hand side: Spherocity separation for fragmentation functions, results for $e^{+}e^{-}$ 
   collisions, the $10\%$ of the most jetty (red points) and isotropic (blue points) events are shown. Right-hand side: It's shown the ratio of each spherocity percentile intervals to jetty events  ($0 < S_{Opc} < 10\%$)  (on top), and the ratio of Pythia 6.4 to Pythia 8.2 results (on bottom). 
   }
  \label{fig:FFSO}
  \end{center}
\end{figure}

In this section, we present the results for $e^{+}e^{-}$ collisions, 
  for spherocity separation in different spherocity percentile intervals.
We observe in  Figure \ref{fig:FFSO} (left)
 that for jetty events ($0 < S_{Opc} < 10\%$) the greater probability for fragmentation is given as $z$ increases. The comparison of the remaining spherocity percentile intervals with respect to the jetty events shows a variation of the fragmentation as the event shape gets isotropic. A crossing point was found as at $z = 0.15$ as shown in Figure  \ref{fig:FFSO}  (right). 
Also, we found that at high $z$ values the difference increases. 

\section{Conclusions}
We have presented an analysis using the anti-$k_{T}$ jet reconstrucction algorithm and also a new method to study the fragmentation functions for jets, based on the event shape called spherocity.
From the Fastjet analysis we obtain that for high $p^{ch jet}_{T}$ the probability for hadron formation
decrease at high $z$ compared to low $p^{ch jet}_{T}$.  
From the spherocity analysis we observe that the greater probability for fragmentation at large $z$ is given for jetty events. 

This analysis could be used as a proposal to study these effects in real data at LHC and in the future FCC and also for tuning high energy Monte Carlo generators.

\section{Aknowledgements}
The authors want to thank the support for this provided from CONACYT and the COST Action CA16201 PARTICLEFACE.
 H.B. want also to thank to CONACYT  for the postdoctoral fellowship and also thank to the Parque de Innovacion Tecnologica (PIT-UAS) for the computational facilities at UAS.

\end{document}